\documentclass[preprint,prl,aps,a4paper]{revtex4-1}
\usepackage{epsfig}

\renewcommand\vec[1]{\mathchoice{{\mathord{\mbox{\boldmath$\displaystyle #1$}}}}{{\mathord{\mbox{\boldmath$\textstyle #1$}}}}{{\mathord{\mbox{\boldmath$\scriptstyle #1$}}}}{{\mathord{\mbox{\boldmath$\scriptscriptstyle #1$}}}}}

\begin{document}

\title{Multiferroics and beyond: electric properties of different magnetic textures}

\author{D.~I.~Khomskii}
%%\author{D. I. Khomskii\\[\medskipamount]
%%\normalsize II. Physikalisches Institut, Universit\"at zu K\"oln,\\[-\smallskipamount]
%%\normalsize Z\"ulpicher Str. 77,\\[-\smallskipamount]
%%\normalsize 50937 K\"oln, Germany}

%%\date{}

%%\maketitle

\affiliation{II. Physikalisches Institut, Universit\"at zu K\"oln, Z\"ulpicher Stra\ss e 77, D-50937 K\"oln, Germany}

\begin{abstract}
This article presents a survey of many nontrivial effects
connected with the coupling of electric and magnetic degrees of
freedom in solids --- the field initiated by I.~E.~Dzyaloshinskii in
1959. I briefly consider the main physics of multiferroic materials,
and concentrate on different effects ``beyond multiferroics'', based on
the same physical mechanisms which operate in multiferroics. In
particular they lead to  nontrivial electric properties of different
magnetic textures --- such as the appearance of dipoles on magnetic
monopoles in spin ice, dipoles on some domain walls in the usual
ferromagnets, on skyrmions etc. The inverse effect, the
appearance of magnetic monopoles on electric charges in
magnetoelectrics, is also discussed. This nontrivial electric
activity of different magnetic textures has manifestations in many
experimental properties of these materials, and it can potentially
lead to novel applications.
\end{abstract}

\maketitle

Electricity and magnetism are two sides of the same physical
phenomenon, which one sees for example from Maxwell equations. Their
strong interplay is crucial for physics, but also for industry. A new
twist in this story is the rapid development of a new field of
spintronics~\cite{Ziese},~\cite{Zutic},
%%Strong coupling of electricity and magnetism is one of the cornerstones of modern physics, and it also plays extremely important role in the industry. It goes back to James Clarke Maxwell with the famous Maxwell equations, and even earlier, to the experiments of Michael Faraday.  Recently it acquired novel significance due to the rapid development of the new field of spintronics,
which uses not only
charge but also spin of electrons in electronic applications.
Modern development of this field can be traced back to the seminal
works of I.~E.~Dzyaloshinskii on weak ferromagnetism, introducing the
concept of antisymmetric exchange --- the Dzyaloshinskii, or
Dzyaloshinskii--Moriya (DM) interaction~\cite{Dzyaloshinskii}, and
almost at the same time suggesting the idea of linear magnetoelectric
effect~\cite{Dzyaloshinskii2}. Both these papers strongly influenced the development of
magnetism. In particular the second paper gave rise to an enormous
development in such fields as magnetoelectrics and multiferroics
(MF)~\cite{Fiebig,Khomskii,Ehrenstein,Cheong,Ramesh,JPCM,Khomskii2,Wang2,Spaldin}.
And the know-how we got in studying these materials we can
apply now to the investigation of many other phenomena dealing with
the coupling of electric and magnetic properties of materials --- and
not only in special magnetoelectric and multiferroic compounds, but
also in ordinary magnetic materials, in particular displaying a
nontrivial electric activity of different magnetic textures:
magnetic domain walls, defects, skyrmions etc. This is the topic of
this paper which has partially a character of a review, but
also contains some novel material.   Thus in the
title of this paper, ``Multiferroics and beyond'' the accent
will be made on the second word, ``beyond''. It is also noteworthy
that in the most recent publications of Dzyaloshinskii he returned
to this field~\cite{Dzyaloshinskii3,Dzyaloshinskii4,Fechner,Meier}.  The name of
Dzyaloshinskii indeed appears in this field over and over again.

\section{Multiferroics; a bit of history}
I will start by very briefly describing the developments of
investigations in the field of multiferroics, as I see it. As
mentioned above, the real activity in this field started after the
publication of the seminal paper by Dzyaloshinskii~\cite{Dzyaloshinskii2},
although people sometimes also cite a short sentence from a much
older paper by Pierre Curie from 1894~\cite{Curie}, in
which he mentioned that in principle one can combine in one material
nontrivial electric and magnetic properties. But this was just a
general declaration, without any specific physics involved. After the
suggestion by Dzyaloshinskii (following a short remark in~\cite{Landau}) that in particular magnetic systems
there may exist magnetoelectric effect --- electric polarization induced by magnetic field,
and the inverse effect --- inducing
magnetization by electric field, this effect was very quickly
discovered in Cr$_2$O$_3$ by Astrov~\cite{Astrov}.
Soon after that a rather rapid development followed.
And people started to look not only at magnetoelectrics, in which one
needs external fields to induce interesting effects, but also at
materials which in the ground state, in the absence of external
fields, would combine the properties of magnets and ferroelectrics. Such
materials were dubbed multiferroics~\cite{Schmid}.
Besides purely scientific interest, these systems
promise very important practical applications --- at present the most
important being probably the potential possibility to control
magnetic memory in computer storage devices electrically, using
effects such as gating, etc.,\ i.e.\ avoiding using electric currents
with their dissipation. This is still the main
driving force of these investigations.

Especially active search and study of such systems at an early
stage of these investigations was carried out in the former Soviet
Union, mainly by two groups; by Smolenskii in Leningrad, now St.~Petersburg,
and by Venevtsev in Moscow, see e.g.~\cite{h1} and~\cite{h2}. In these two groups a
number of multiferroics were discovered, but the practical use of
these was restricted either by low temperatures at which such state
existed or by relatively weak coupling between electric and magnetic
subsystems --- which even now
remain the main obstacles to their wide practical use, although
enormous progress has been achieved in recent years. But, besides the
potential applications, the study of these materials presented some real
challenges to the general physics. One of these was an early
observation that in one of the biggest classes of materials ---
perovskites $AB$O$_3$, to which quite a lot of magnetic system belong,
including the famous colossal magnetoresistance manganites such as
La$_{1-x}$Sr$_x$MnO$_3$, but also most of the interesting and practically
important ferroelectrics, starting from BaTiO$_3$, there exists a striking
``mutual exclusion'': materials with transition metals $B$ with
partially-filled $d$-shells are magnetic, whereas those with empty
$d$-shell, with configuration $d^0$, are of course not magnetic but could
become ferroelectric. And surprisingly there was practically no
overlap between these two big classes of materials: either these were
ferroelectric, or magnetic, but almost never both simultaneously. This $d^n$--$d^0$
dichotomy was noticed long ago, but did not attract attention for
quite a while. I remember that at around 1996 I told this
story, and in general of the attempts to combine in one material both
(ferro)magnetic and ferroelectric properties, to a very good and
extremely broad physicist George Sawatzky with whom we have both worked
at that time in Groningen University, and his reaction was very
characteristic: ``But this is very interesting! Why don't we know
anything about it?'' Couple of years later at the workshop on quantum
magnetism in KITP in Santa Barbara in 1998 we organized the
brainstorming discussion of the topic of a possible coexistence of
electricity and magnetism. Nicola Spaldin (at that time Nicola Hill)
spoke at this meeting about her {\it ab initio} calculations of one such
``suspicious'' material, BiNiO$_3$~\cite{Hill}, and we discussed
this empirical observation about mutual exclusion of ferroelectricity
(with $d^0$ configuration) and magnetism ($d^n$ configuration).

The next
important step happened in 2001 when Nicola organized a special session
at the March meeting of American Physical Society, Session C21,
devoted to the discussion of multiferroics~\cite{BAMPS}. One can say that this
session actually has lead to the revival of interest in this
problem, and ``put multiferroics on the map.'' And indeed, already in 2007 there were 7 special
sessions on multiferroics at the APS March meeting --- sessions, not
talks! And in 2008 the term ``multiferroics'' was in the title of 12
sessions at the March meeting.  And apparently the start was this first
special session at the March meeting in 2001.

But of course the most important was the experimental progress in
this field reached a bit later, mainly by three groups. Tsuyoshi
Kimura and Yoshinori Tokura found striking MF behaviour in TbMnO$_3$~\cite{Kimura},
and Sang-Wook Cheong with
coworkers in Tb$_2$Mn$_2$O$_5$~\cite{Hur}. This
was actually the discovery of a novel type of multiferroics which one now
calls type-II MF --- the systems in which ferroelectricity appears
and is driven by a particular type of magnetic ordering, in contrast
to type-I multiferroics in which FE and magnetism appear
independently and in which most often different subsystems and different
ions are responsible for the two orderings. The third
breakthrough was the synthesis by the group of Ramesh of thin films
of the classical type-I multiferroic BiFeO$_3$ (BFO)~\cite{Wang}, till now remaining the system with the
best performance and probably the best perspectives for practical
applications (if we speak of one material, not of composite systems
such as e.g.\ multilayers of good FE and good ferromagnets).  The films
made by Ramesh have shown very spectacular properties, with much
stronger effects than known until that time on the bulk BiFeO$_3$
(although now people reach such good performance also in
the bulk BFO)\null. These three experimental breakthroughs, together with
the realization of some fundamental  theoretical problems and
challenges, have lead to the revival of common interest in
multiferroics and to a rapid progress in this field. At present quite
a lot of new MF systems are found, and we can probably say that the
main physical mechanisms responsible for this phenomenon are already
understood, although the constant progress in this field is still
ongoing and novel materials and novel
phenomena are found. And, besides multiferroics per se, the
experience and know-how we have gained by
studying multiferroics can be applied to the discussion of related
phenomena even in non-MF materials. In this paper I will try to
summarise some of this novel development, although of course I
cannot cover this very big field. There exist already quite a few
general review articles on multiferroics~\cite{Fiebig,Khomskii,Ehrenstein,Cheong,Ramesh,JPCM,Khomskii2,Wang2,Spaldin},
a section on multiferroics is included in the book~\cite{Khomskii-book}, and a special book on these systems is
published~\cite{wang-wang}. I will not discuss here these
effects in details, but will rather concentrate on the ``spin-offs'' of
these studies, on a relatively qualitative level, concentrating on
what is going on ``beyond multiferroics''.  But I will start with a
short synopsis of the main effects and mechanisms of multiferroics per~se.

\section{Multiferroics: basic physics}

Generally speaking, we can divide multiferroics into two big groups.
The first one, which can be called type-I MF~\cite{Khomskii2}, is formed by materials
in which magnetism and ferroelectricity appear
independently and are due to different mechanisms, different
subsystems in a material --- although of course they are coupled. In
these materials the values of critical temperatures of magnetic
and FE transitions can often be quite high, with the FE transition
occurring typically at higher temperatures. The best example of these type-I
MF is the already-mentioned BiFeO$_3$ with $T_{\rm FE} =1100\,\rm K$
and $T_{\rm N} = 643 \,\rm K$, or the hexagonal manganites $R$MnO$_3$  ($R$ is a rare earth) with $T_{\rm FE} \sim 1000\,\rm K$
and $T_{\rm N} \sim 100\,\rm K$. The magnetic and FE degrees of freedom in these systems
are of course coupled but this coupling is typically rather weak.

Another group of multiferroics, type-II MF, are those in which
ferroelectricity is driven by a particular type of magnetic ordering.
To these belong the two multiferroics of this class discovered first, TbMnO$_3$~\cite{Kimura} and TbMn$_2$O$_5$~\cite{Hur}.
The paramagnetic state in these systems is not FE, but
a particular type of magnetic ordering can generate FE polarization. It
is these new materials which have attracted the main general
physical interest and which have lead to the emergence
of several novel physical concepts.  From a
practical point of view these type-II MF may seem more promising ---
the intrinsic coupling of magnetism and ferroelectricity in those is
of course very strong. But unfortunately most of these materials have
relatively low values of critical temperatures, below which this
coexistence of FE and magnetism is observed.

\begin{figure}
\includegraphics{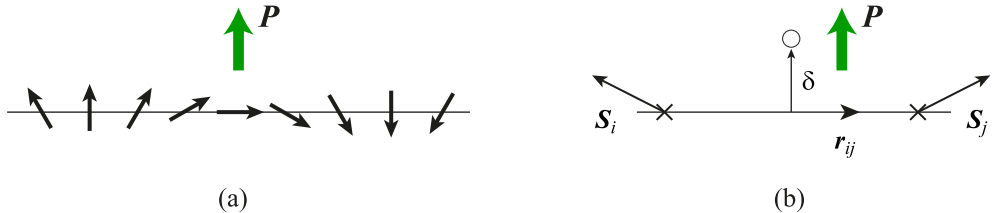}
\caption{\label{FIG:1}(a)~Cycloidal magnetic structure and the resulting electric polarization.
(b)~The mechanism (inverse Dzyaloshinskii mechanism) of the appearance of the dipole moment,
or polarization, for a pair of spins, cf.\ Eq.~(\ref{eq:P-polar}). Here crosses
are positively charged TM ions and the circle is e.g.\ O$^{-2}$ ion. The thick green
arrow is the electric dipole moment or electric polarization.}
\end{figure}

Two general mechanisms are involved to explain the appearance of
electric polarization in some particular magnetically-ordered states
in type-II multiferroics. One of those is the usual
magnetostriction: a particular magnetic ordering can break inversion
symmetry, and the corresponding magnetostrictive distortion of the
lattice in some magnetic structures can induce electric polarization,
see e.g.~\cite{vdBrink}.  This mechanism does not require the presence of
spin--orbit interaction.  Another, more widespread and more
interesting mechanism of MF behaviour in type-II MF relies on the
use of relativistic spin--orbit coupling and is more in spirit with the
original mechanism of magnetoelectricity proposed by Dzyaloshinskii.
There are several versions of this mechanism, see e.g.~\cite{Yoshinori,Ono}.
The most popular and most important for going ``beyond
multiferroics'' is the mechanism of the appearance of electric
polarization in magnets with cycloidal magnetic structure. This
mechanism was elucidated in a microscopic treatment in the paper of
Katsura, Nagaosa and Balatsky~\cite{Katsura}, and it was obtained
using the Landau expansion by Mostovoy~\cite{Mostovoy}.  According to this theory, if spins on two
neighbouring magnetic ions are not collinear, there would appear for
this pair of ions an electric polarization  proportional to
\begin{equation}
\vec P_{ij}  = c\vec r_{ij} \times [\vec S_i \times \vec S_j]
\label{eq:P-polar}
\end{equation}
where $c$ is some coefficient. For cycloidal magnetic structure shown
in Fig.~\ref{FIG:1}(a), it would give a net polarization, i.e.\ the cycloidal
magnets are intrinsically multiferroic.

The microscopic mechanism of the appearance of electric polarization
was disclosed by Sergienko, Sen and Dagotto~\cite{Sergienko}.  It is
actually the inverse Dzyaloshinskii effect.  According to~\cite{Dzyaloshinskii}, for some particular symmetries
there exists for a pair of ions $i,j$ the antisymmetric DM
(Dzyaloshinskii--Moriya) exchange
\begin{equation}
H_{\rm DM} = -\vec D_{ij} \cdot (\vec S_i \times \vec S_j)\;.  \label{eq:1}
\label{eq:HDM}
\end{equation}
If the Dzyaloshinskii vector $\vec D$ is nonzero, this interaction leads to
canting of neighbouring spins. But vice versa, if spins are
noncollinear for any reason, to gain this energy it may be
favourable to distort the lattice and shift the ions so as to make
$\vec D \neq \vec 0$.  In typical cases, e.g.\ in perovskites, the nn
exchange between magnetic ions $M_i$ and $M_j$ is carried out by
superexchange via ligand (e.g.\ oxygen ions) sitting in between, Fig.~\ref{FIG:1}(b).
If such oxygen sits exactly at the middle of the ($ij$) bond, by
symmetry the DM interaction is zero~\cite{Moriya}. To gain the DM energy (\ref{eq:HDM}) it is then better to shift such
oxygen by some distance $\vec\delta$ perpendicular to the ($ij$) bond, e.g.\ in
the $z$-direction, Fig.~\ref{FIG:1}(b). Then there would appear a nonzero DM
interaction with the vector $\vec D \sim \vec r_{ij} \times \vec\delta$,
so that now we gain the  DM energy~(\ref{eq:HDM}). But such shift of negatively
charged oxygen ions away from the ``centre of gravity'' of positive
charges of TM ions $M_i$ and $M_j$ would create an electric dipole, or
a polarization in the $z$-direction. This is the inverse Dzyaloshinskii
effect --- the mechanism of the appearance of MF behaviour in many MF
systems, and it is this mechanism which could also lead to an
electric polarization of different magnetic textures --- some domain
walls, skyrmions, etc., which I will extensively ``exploit''  below.

\section{Electronic mechanisms of coupling of electricity and magnetism}
\begin{figure}
\includegraphics{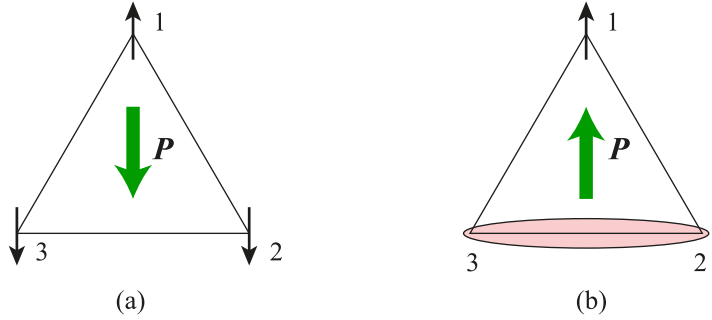}
\caption{\label{FIG:2}The appearance of electric polarization in triangles with a particular spin pattern.
The oval in~(b) is the singlet state of spins $\vec S_2$, $\vec S_3$, with $\vec S_2 + \vec S_3 = \vec 0$
and $\vec S_2\cdot \vec S_3 = -\frac34$, cf.\ Eq.~(\ref{eq:n1}).}
\end{figure}

Usually the appearance of electric polarization in magnetoelectrics
and multiferroics is attributed to lattice effects, to the
corresponding shifts of ions. However there also exists a purely
electronic mechanism which can lead to such coupling. In particular
this mechanism can operate in frustrated systems, the basic building blocks of
which are triangles made of transition metal (TM) ions, Fig.~\ref{FIG:2}. If one
describes $d$-electrons in such a triangle by the usual Hubbard model,
\begin{equation}
H = -t\sum_{\langle ij\rangle,\sigma}c^\dagger_{i\sigma}c^{\vphantom{\dagger}}_{j\sigma}    +   U\,\sum_i n_{i\uparrow}n_{i\downarrow}  \;,  \label{eq:MODEL}
\end{equation}
one can show that for certain magnetic textures, for certain spin
correlation functions, there may appear charge
redistribution in such triangles, so that in contrast to the standard picture the charge at
site~$i$ is not exactly~1 as in the usual Mott insulators, but can be more
or less than~1. Correspondingly, there would appear in this situation
an electric dipole in such a triangle, the magnitude and direction of
which is determined by the spin structure. That is, it is a purely electronic
mechanism of multiferroic behaviour.

The treatment carried out in~\cite{Bulaevskii} (see also~\cite{KhomskiiJPCM})
shows that when one
treats the model (\ref{eq:MODEL}) (for one electron per site and for strong
correlations $t/U \ll 1$) to third order in $(t/U)$, the charge at
site~1 is given by
\begin{equation}
n_1 \sim 1 + K \, [\vec S_1\cdot (\vec S_2 + \vec S_3) - 2 \vec S_2 \cdot \vec S_3]\;, \qquad  K=8t^3/U^2   \label{eq:n1}
\end{equation}
and similar expressions for $n_2$, $n_3$.  I.e.\ if this
spin correlation function entering (\ref{eq:n1}) is nonzero, there would occur
charge redistribution in such a triangle, and there would appear in it
an electric dipole moment, see Fig~\ref{FIG:2}.  It seems that the effect here
is rather small, $\sim t^3/U^2 \ll 1$. But one can show that the same
expression would also be valid in case when $t \sim U$, only with a different
value for  the coefficient $K$ in~(\ref{eq:n1}). Thus in general this purely
electronic mechanism of the generation of electronic polarization by
a particular magnetic texture may be quite significant. (Actually
the contribution of lattice distortions would lead to the same
expression for polarization in this case, with the same dependence
on the spin structure~\cite{KhomskiiJPCM}.)

Interestingly enough, in the same situation there would appear for
certain spin textures also spontaneous circular currents in such
triangles, with the respective orbital moment. These currents exist
for noncoplanar spins, and they are given by the expression~\cite{Bulaevskii}
\begin{equation}
j_{123} = C\kappa(123)
\label{eq:j123}
\end{equation}
where $\kappa$ is the scalar spin chirality,
\begin{equation}
\kappa(123) = \vec S_1 \cdot (\vec S_2 \times \vec S_3)
\label{eq:kappa123}
\end{equation}
Here the coefficient $C$ in the nondegenerate Hubbard model is given by
$C=24et^3/hU^2$. Thus such noncoplanar spin texture
will lead not only to the fictitious magnetic field due to Berry
phase~\cite{Taguchi}, but also to real orbital currents and
orbital moments, proportional to scalar spin chirality~(\ref{eq:kappa123}). We will see the examples of this
effect below.

The physical mechanisms described in this and in the previous sections
can be used to predict or explain not only some features of
multiferroics,  but also some phenomena connected with the interplay
between electric and magnetic degrees of freedom in other situations.
This will be discussed in the following sections.

\section{Dipoles on monopoles in spin ice}
\begin{figure}
\includegraphics{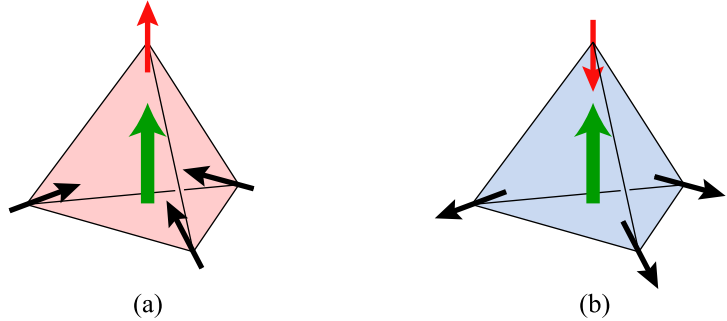}
\caption{\label{FIG:3}The appearance of electric dipoles (thick green arrows) for monopoles~(a) and antimonopoles~(b)
in metal tetrahedra --- the building blocks of spin ice pyrochlores.}
\end{figure}

One very interesting phenomenon, discovered recently in frustrated
systems, is the appearance in some of them (so called spin ice systems) of
excitations having the properties of magnetic monopoles~\cite{Ryzhkin,Castelnovo}.
After this proposal was made theoretically,
such monopoles where discovered and extensively studied
experimentally, see e.g.~\cite{Morris,Fennel,Mengotti}.
Such monopoles were first predicted and observed in pyrochlore spin
ice such as Tb$_2$Ti$_2$O$_7$, in which the main building blocks are metal (here
Tb) tetrahedra with strong Ising ions, with moments pointing towards
(or away from) the centre of tetrahedra, Fig.~\ref{FIG:3}. In the usual spin
ice ground state we have the situation with (2-in)--(2-out) state at
each tetrahedron; such spin distribution is not unique and is highly
frustrated. The monopole configuration which for the usual spin ice
is an excited state, which can be excited at finite temperature
but which can also be stabilised by the  external magnetic field,
corresponds to the (3-in)--(1-out) state (monopole $\mu$ with the magnetic
charge inside a tetrahedron~$+2Q$, where each spin is represented by a pair of magnetic
charges $(+Q,-Q)$), or (1-in)--(3-out) state
(antimonopole $\bar\mu$, with the charge~$-2Q$). Such monopoles and
antimonopoles can move in pyrochlore spin ice by flipping some spins and leaving a trail of
reversed spins, but due to spin disorder inherent in spin ice state
such strings have no tension (the energy does not increase linearly
with the length of the string, as it would in a long-range ordered
state), i.e.\ there is no confinement, and such monopoles and antimonopoles can live in a crystal
as independent excitations.

This concerns the magnetic degrees of freedom of these systems.
But using the treatment presented in the previous section, see
Eq.~(\ref{eq:n1}), one can show that there would exist in this case electric
dipoles attached to each magnetic monopoles~\cite{Khomskii-ice}. Indeed, using Eq.~(\ref{eq:n1}) one can see
that there would be no net dipoles for the standard (2-in)--(2-out) states
of spin ice (and neither for (4-in) or (4-out) states). But from this
expression we immediately see that there would be a nonzero dipole
(thick green arrow) in monopoles and antimonopoles, directed towards
the ``special'' spin (red spin in Fig~\ref{FIG:3} --- the out-spin in
monopole, in-spin in antimonopole). (As Eq.~(\ref{eq:n1}) is even in spins, the
reversal of all spins in going from monopole to antimonopole leaves
the dipole unchanged.) Such electric dipoles at each magnetic
monopole in the usual spin ice would be random and dynamic, and they
would give extra electric activity in the state with monopoles.
These dipoles and their consequences were indeed observed
experimentally in Dy$_2$Ti$_2$O$_7$ and Tb$_2$TiO$_7$~\cite{Grams,Jin}.
In a strong magnetic field $\vec H \parallel [111]$ there would
appear ordered monopolies and antimonopoles at every site in spin
ice pyrochlore, and correspondingly there would be also dipole
moments at every tetrahedron, ordered in an antiferroelectric
fashion~\cite{Khomskii-ice}.

\begin{figure}
\includegraphics{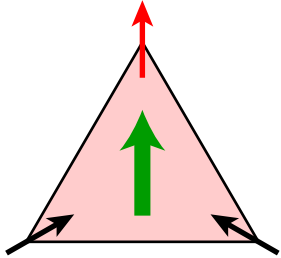}
\caption{\label{FIG:4}The appearance of an electric dipole of a magnetic triangle --- the building block
of kagome spin ice systems.}
\end{figure}

A similar effect should also exist in kagome spin ice, Fig.~\ref{FIG:4}. In
contrast to pyrochlores, here monopole ((2-in)--(1-out)) or
antimonopole ((1-in)--(2-out)) configurations would exist at each
triangle already in the ground state.  Consequently there would
exist, according to Eq.~(\ref{eq:n1}), electric dipoles at every triangle.
Again, in the real spin ice ground state the spin configurations, i.e.\ in
this case monopoles and antimonopoles, would be random, and with
them the dipoles would also be random and fluctuating.

\begin{figure}
\includegraphics{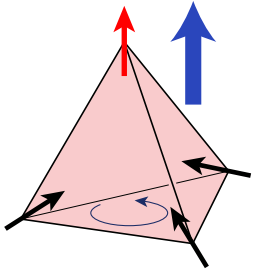}
\caption{\label{FIG:4'}Spontaneous current (thin circular arrow) and the resulting orbital moment (thick blue arrow)
at a magnetic monopole in pyrochlore spin ice.
As the expression~(\ref{eq:kappa123}) is odd in spins, currents and orbital moments in antimonopoles are
opposite to those on monopoles, in contrast to electric dipoles}
\end{figure}

Using Eqs.~(\ref{eq:j123}),~(\ref{eq:kappa123}) one can show that there
also exist spontaneous currents and orbital moments at magnetic monopoles in pyrochlores spin ice,
Fig~\ref{FIG:4'}. These are random and dynamic in spin ice state with excited monopoles.
(Note that, in contrast to dipoles, currents also exist in the usual (2-in)--(2-out) tetrahedra.)

\section{Moment fragmentation and dipoles}
\begin{figure}
\includegraphics{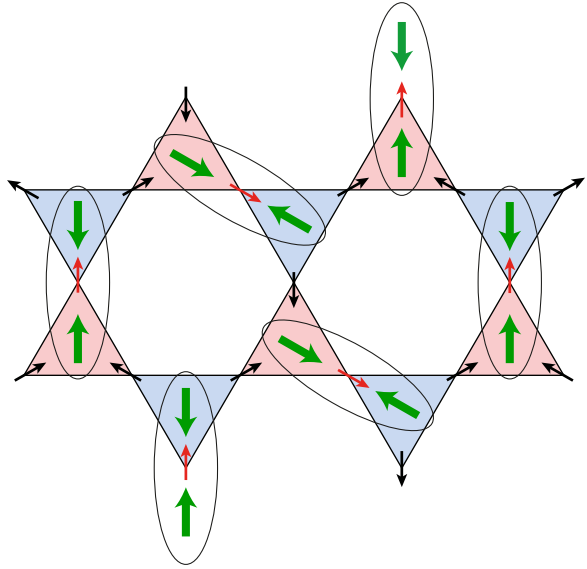}
\caption{\label{FIG:5}A kagome spin ice with moment fragmentation. In pyrochlore
spin ice with moment fragmentation such state looks similar~\cite{Brooks-Bartlett,ya-arXiv}.
Electric dipoles (thick green arrows) are also shown.}
\end{figure}

An interesting twist in this story is the recent suggestion~\cite{Brooks-Bartlett}
that there could exist in spin ice systems a
novel state --- the state with spin, or magnetic moment fragmentation.
This idea relies on the Helmholtz decomposition of magnetization
into a divergence-free and divergence-full components, the later
actually describing the distribution of magnetic monopoles --- the
sources of magnetic field in a system.

\begin{equation}
\vec M(\vec r) = \vec M_{\rm mono}   +  \vec M_{\rm ice}   =  -\nabla \rho  + \mathop{\rm curl} \vec A
\label{eq:deco}
\end{equation}
where $\rho(\vec r)$ is the density of magnetic charges (``monopoles''), and the
second part is the divergence-free part of magnetization, corresponding
to pure spin-ice configuration (with (2-in)--(2-out) states at every
tetrahedron, with  zero total magnetic charge inside every
tetrahedron).  (The decomposition (\ref{eq:deco}) of course does not violate
the Maxwell equations, in particular $\mathop{\rm div}\vec B=0$ as $\vec B= \vec H + 4 \pi \vec M$,
and $\mathop{\rm div}\vec H = -4 \pi \mathop{\rm div} \vec M$, see e.g.~\cite{Bramwell}.)
Here the most interesting feature is that there
may exist a nontrivial partially-ordered state: the state in which
monopoles and antimonopoles exist and are perfectly ordered in the
ground state, whereas spins themselves are still disordered, see Fig.\ref{FIG:5}
for the kagome spin ice with moment fragmentation~\cite{Moller,Chern}.

A question arises, what would become with electric activity with
dipoles in such a state. One can show~\cite{ya-arXiv} that dipoles would
still exist in such states, but they would not be free independent
dipoles but, rather, they would always be paired into $(\vec d, -\vec d)$
pairs, Fig.~\ref{FIG:5}. That is, the transition to such moment fragmentation
state would lead to the reduction of electric activity, e.g.\
microwave absorption, in such a state.  The spontaneous currents and the corresponding
orbital moments on monopoles are also paired, in this case in $(\vec L,\vec L)$ pairs
with parallel orbital moments~$\vec L$~\cite{ya-arXiv}.

\begin{figure}
\includegraphics{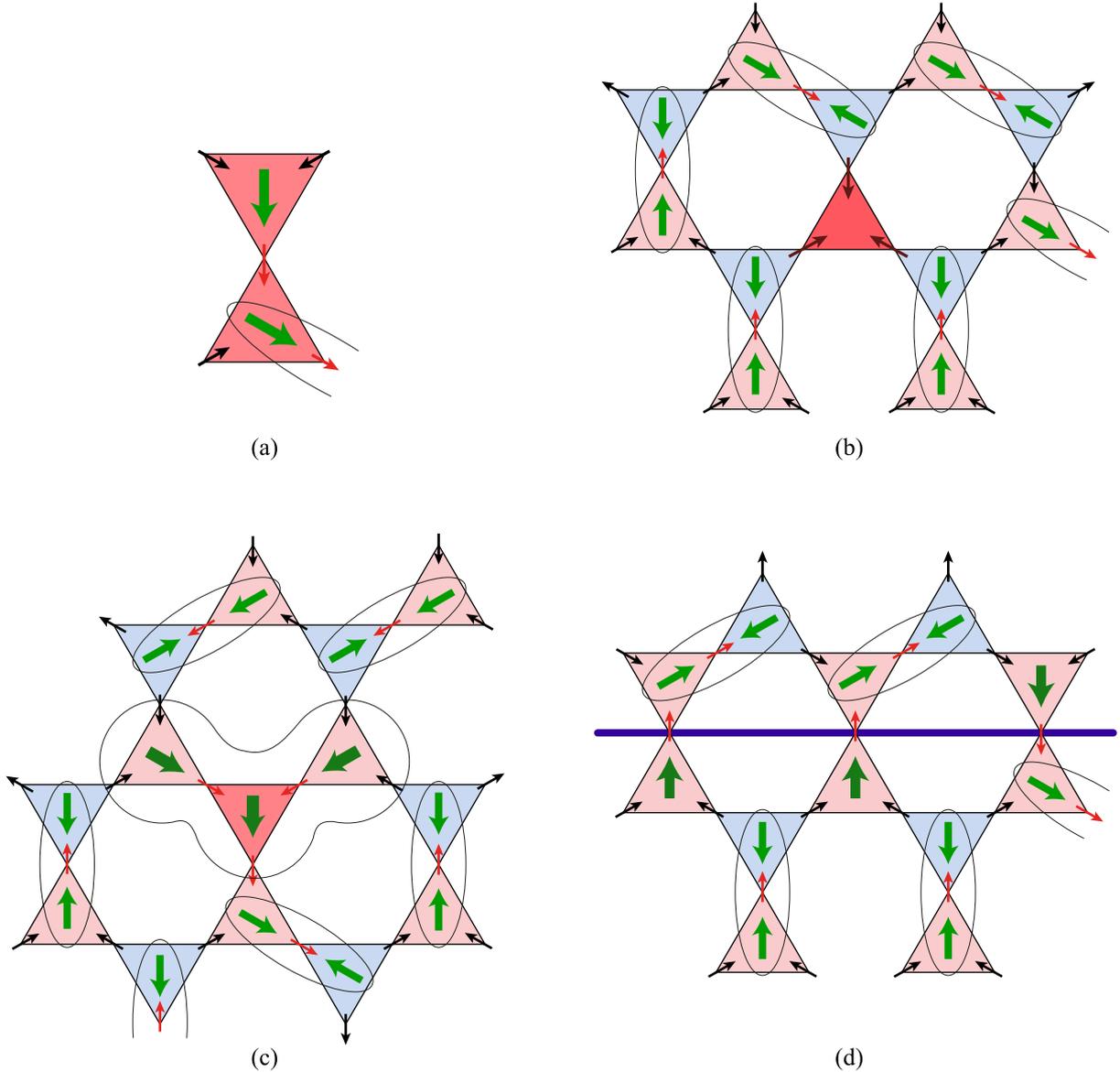}
\caption{\label{FIG:6}Electric dipoles on defects in moment fragmentation state in kagome spin ice
(the situation in pyrochlores is similar~\cite{ya-arXiv}).
(a)~A typical situation for defects and domain walls: two neighbouring triangles both with monopoles,
showing that at least one of them would have an unpaired dipole.
(b)~The ``supermonopole'' defect (3-in triangle). By recommuting spins one can get rid of unpaired dipoles.
(c)~The appearance of unpaired dipoles at a monopole in place of antimonopole.
(d)~One type of domain wall in monopole ordered state, showing the appearance of unpaired dipoles on it.}
\end{figure}

A very special feature of moment fragmentation state in spin ice is the coexistence
in one spin system of both long-range ordered (monopoles) and disordered (spin themselves) components.
This unusual situation determines also the unusual properties of defects and domain walls
in these systems.
Each time
we have an ordered state we have to think of defects or excitations
breaking this perfect order, and also of creation of domains and
domain walls~\cite{Khomskii-book1}. The situation is of course the same here.
We can
create different types of point defects.  These are for example the ``reversed'' monopole, i.e.\
monopole in place of antimonopole, or novel excitations, such as the
state with (3-in) (``supermonopole'', magnetic charge $3Q$ inside a
triangle) state in kagome ice, in which there are monopoles and
antimonopoles (charges~$\pm Q$) at every triangle, ordered in moment fragmentation
state. One can also form domain walls in $\mu-\bar\mu$ two-sublattice
ordered state (such as domain wall in an antiferromagnet). And all these
defects or textures in the ordered monopole sector coexist with the
still preserved spin disorder!

One can show that such defects or domain walls would also modify
electric properties. At first glance it may seem that every such
defect would lead to the creation of free dipoles: dipoles are paired
into $(\vec d, -\vec d)$ pairs in the fragmentation state, and defects
would remove one dipole for the pair, leaving another dipole
unpaired. But actually it is not always the case. Thus the
``supermonopole'' ((3-in) state in kagome ice, or the (4-in) tetrahedron in
a pyrochlore) would not create such free dipoles: using the remaining
freedom of spins one can ``recommute'' those in such way that free
dipoles are removed, Fig.~\ref{FIG:6}(b). But other types of defects, e.g.\ $\mu$ in
place of $\bar\mu$, would necessarily lead to the creation of unpaired
dipoles (even three of them in kagome ice with spin fragmentation,
Fig.~\ref{FIG:6}(c) and four in moment fragmentation pyrochlores).
A typical configuration of such defects is the neighbouring pair of two monopoles, see Fig.~\ref{FIG:6}(a).
One sees that in such situation there would be at least one unpaired dipole.
This leads to the unpaired dipoles at the defect in Fig.~\ref{FIG:6}(b).
Also different types of domain
walls would have unpaired dipoles, see e.g.~Fig.~\ref{FIG:6}(d). This
probably can be used to control, orient and move such defects or
domain walls by (inhomogeneous) electric field, which could
potentially be useful for some applications.

\section{Dipoles at magnetic textures in ordinary magnets: domain walls, skyrmions and all that}
\begin{figure}
\includegraphics{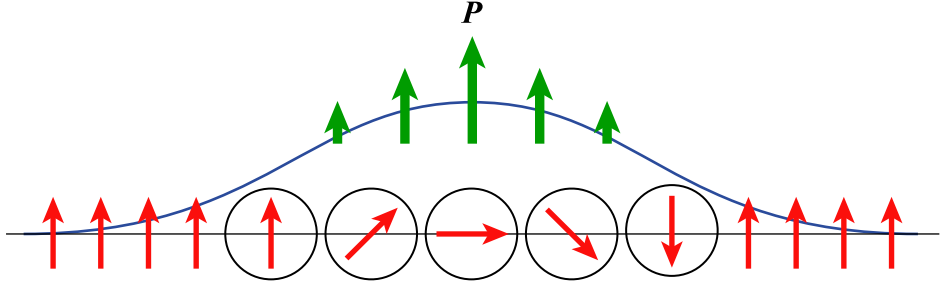}
\caption{\label{FIG:7}Electric dipoles at the N\'eel domain wall in a ferromagnet.}
\end{figure}

\begin{figure}
\includegraphics{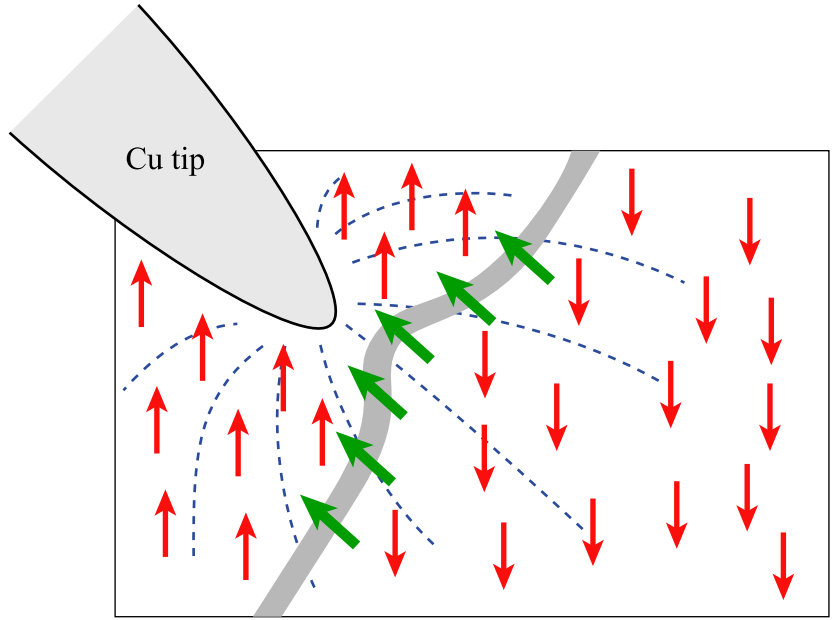}
\caption{\label{FIG:8}The scheme of the experiment of Ref.~\cite{Logginov} illustrating
the motion of the N\'eel domain wall in an ordinary insulating ferromagnet
when a voltage pulse is applied to a metallic tip, creating an inhomogeneous
electric field in the sample. Red arrows are spins, thick green arrows are electric
dipoles at the domain wall.}
\end{figure}

We have seen in treating multiferroics that there should appear
electric dipoles or polarization for certain spin configurations,
e.g.\ at spin cycloids, Fig.~\ref{FIG:1}(a). But the same local spin
configuration can exist in many other situation, e.g.\ at the N\'eel
domain walls in ordinary ferromagnets, Fig.~\ref{FIG:7}.  The magnetic structure of such domain wall can be
visualised as part of a cycloid, i.e.\ according to the expression (\ref{eq:P-polar})
it should also have nonzero electric polarization~\cite{Mostovoy}.  Consequently one could think of
influencing such domain walls by electric field. This idea was
proposed by Dzyaloshinskii in~\cite{Dzyaloshinskii3}. But already before that the
team in Moscow University had the same idea, and they carried out
corresponding experiments~\cite{Logginov}. Using films of the
standard magnetic garnet --- a good insulating ferromagnet with $T_c$
above room temperature --- the authors of~\cite{Logginov} of  managed to see the motion of N\'eel domain
walls when they applied an inhomogeneous electric
field to the sample (created simply by applying a voltage pulse to a sharpened copper
wire close to the film), see Fig.~\ref{FIG:8}: the electric dipoles existing at N\'eel domain walls
were attracted to the region of stronger electric field close to the tip.
This is indeed a conceptually very simple but beautiful experiment
confirming the main physical ideas developed first in treating
multiferroics but which can be now applied to many other situations.

Yet another experiment which can be explained by the same physics ---
the coupling of magnetic and electric degrees of freedom at
particular magnetic textures --- is the observation of the creation of
spiral magnetic structures at thin magnetic layers on the surface
of nonmagnetic metals. It was discovered by the group of Wiesendanger
in Hamburg that when one makes a monolayer of Mn on top of W,
the magnetic structure of Mn becomes actually cycloidal instead of the
expected collinear structure~\cite{Bode}. This effect was explained in~\cite{Ferriani} using microscopic treatment
with the inclusion of the Dzyaloshinskii--Moriya interaction which should
be present at the surface (the surface, of course, breaks inversion
symmetry). But this effect can be explained very simply in the same
picture of the appearance of electric polarization in cycloidal
structures. As is shown in Fig.~\ref{FIG:1}(a) the cycloidal structure leads to
the creation of electric polarization directed in the plane of the
cycloid and perpendicular to its direction. But, inversely, if there
exists in a system an intrinsic polarization, or intrinsic electric
field, it would lead to the creation of a cycloidal magnetic structure.
Such electric field always exists at the surface of a metal --- there
exists there a double layer, or potential drop (the work function) and
the corresponding electric field perpendicular to the layer.
Consequently one may expect that instead of a collinear magnetic
structure there would appear in this case a cycloidal structure with
a particular sense of spin rotations. This is exactly what was observed
in~\cite{Bode}.
Such spin rotation works against spin anisotropy, thus if there
exists too strong anisotropy in the film, e.g.\ an easy plane (which is often
the case in magnetic films), this anisotropy can suppress the creation
of such cycloids. But if such anisotropy is not too strong, cycloids
can form in such situations.  And, besides magnetic spirals, one can also
stabilize magnetic skyrmions in similar situations. This was observed
by the same group in a double layer of Fe on Ir~\cite{Sassmannhausen}.

\begin{figure}
\includegraphics{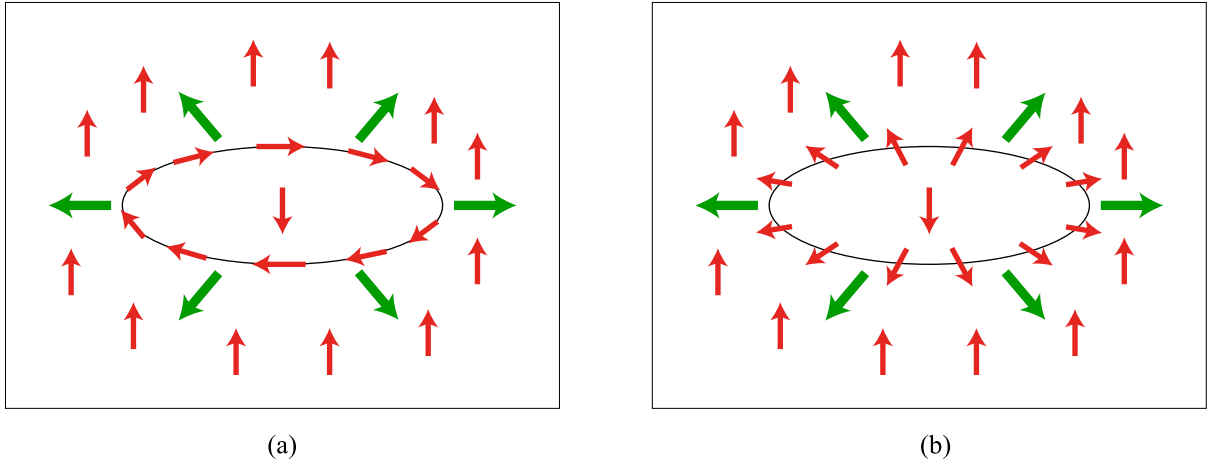}
\caption{\label{FIG:9}Electric dipoles at skyrmions.   Spins are assumed to point up in the bulk and down at the centre of the skyrmion.
In approaching the centre the spins (red arrows) are rotating, and we show the spin direction at the ``middle''
ring where the spins lie in the $xy$-plane.
(a)~Bloch skyrmion.
(b)~N\'eel skyrmion. In both cases, according to Eq.~(\ref{eq:P-polar}), there would exist electric dipoles directed radially (thick green arrows).}
\end{figure}

Yet another interesting effect, connected with the same physics, was
observed by the Hamburg group when they applied an
inhomogeneous electric field to a system. In spirit this experiment is similar to
the experiment of the MSU group described above. The authors of~\cite{Hsu} (see also~\cite{Roch})
found that they can
create skyrmions under the tip with electric voltage. And such
skyrmions were created for one polarity of the field, but not for the
other. Again, at least qualitatively one can explain this observation
by the same physics as described above.  There exist two types of
skyrmions: the ones in which spins are rotating as in Bloch domain
walls, so that in the middle of the skyrmion the spins are oriented along the
circle, Fig.~\ref{FIG:9}(a), and N\'eel skyrmions, in which the spins at the ``middle''
circle point away (or towards) the centre, Fig.~\ref{FIG:9}(b). Using the
expression (\ref{eq:P-polar}) one can show~\cite{Delaney} that there
would appear local electric dipoles in such textures: in both cases
the dipoles will be directed radially, the green arrows in Fig.~\ref{FIG:9}.
(For N\'eel skyrmions there would also appear net polarization
perpendicular to the skyrmion plane.) When we
apply voltage to the tip, we create an inhomogeneous electric
field. Its radial component would interact with the radial dipoles at
the skyrmions and, depending on the polarity, it would lead either to
energy gain or to energy loss. Therefore one could indeed expect that
the field of one polarity would stabilize the skyrmions under the tip,
and the opposite polarity would work against it. This simple picture
can thus explain experimental observation of~\cite{Sassmannhausen}.

\section{Dipoles at spin waves}
The same picture as used in the previous section can also predict a
nontrivial effect even for ordinary spin waves in ferromagnets. Long
ago Bogolyubov posed the question~\cite{Bogolyubov}: is there any nontrivial electric
effect connected with spin waves? He actually considered what later on became known as the
Hubbard model and treated it in perturbation theory, anticipating in
particular the much later treatment of superexchange by Anderson,
Goodenough and others. And he asked a question whether (when treated
not in a purely magnetic models but going back to original
electronic description) magnons can carry small electric charge. In a
sense the treatment in~\cite{Bulaevskii} is in spirit similar to this old
approach of Bogolyubov. His conclusion was somewhat ambiguous:
actually he did not obtain any real current carried out by magnons, but
the expressions equivalent to the currents due to spin
chirality were in fact contained in his results. Recently Morimoto
and Nagaosa~\cite{Morimoto}
addressed the same question but for a specific situation with shift
currents in multiferroics, and obtained that in particular situations
there may indeed exist nontrivial electric effects on magnons.

\begin{figure}
\includegraphics{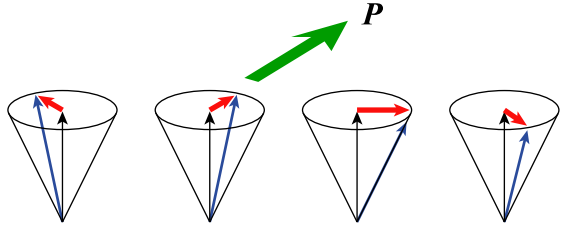}
\caption{\label{FIG:10}The appearance of an electric dipole at the magnon in a ferromagnet.}
\end{figure}

Using the physics described above, in particular the same expression
(\ref{eq:P-polar}) for the electric polarization for canted spins, one can give the
arguments that the ordinary spin waves e.g.\ in ferromagnets will
have electric activity, however they will
carry not electric charge, but electric dipoles~\cite{Khomskii2}.
Indeed the quasiclassical picture of a magnon is that shown
in Fig.~\ref{FIG:9}: in a magnon the spins are somewhat tilted away from the
average magnetization direction~$z$ and precess around it, so that the
snapshot is that shown in Fig.~\ref{FIG:10}. And in a spin wave this spin
structure ``moves'' with a certain velocity.

When we look at the structure of Fig.~\ref{FIG:10} we immediately realize that
whereas the $z$-component of magnetization is constant (and is just
slightly less than $M_{\rm max}$), the perpendicular $xy$-components of
magnetization form exactly the cycloid of Fig.~\ref{FIG:1}! Therefore, by the
same expression (\ref{eq:P-polar}) we should expect that there would appear electric
dipoles at the usual spin waves, shown in Fig.~\ref{FIG:10}. And if we make a
spin wave packet, then when it moves in a sample it would carry with
it not only magnetization, but also a perpendicular electric
dipole. I do not have a good idea could be the experimental consequences
of the presence of such dipoles; maybe they can contribute, for
example, to Raman scattering on magnons, changing selection rules, or
could lead to some other similar effects.

\section{Monopoles on charges in magnetoelectrics}

\begin{figure}
\includegraphics{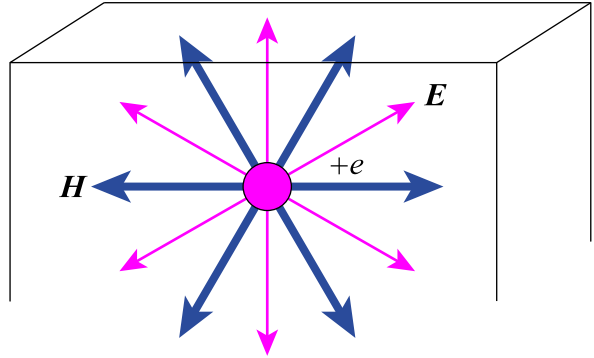}
\caption{\label{FIG:11}The appearance of magnetic monopole (radial magnetic field) at a charge
in a magnetoelectric material (with nonzero diagonal components of the magnetoelectric tensor~$\alpha_{ij}$).}
\end{figure}

In the previous sections we have discussed several situations in which
particular magnetic textures generate electric dipoles or currents.
There is however also an opposite effect --- in certain cases electric
charges can generate magnetic response, and in particular lead to
the creation of magnetic monopoles~\cite{Khomskii-MMUT},~\cite{Fechner}.   This is  the
situation in magnetoelectrics --- the field initiated by
Dzyaloshinskii in 1959.  Indeed if we have a magnetoelectric with
diagonal ME tensor $\alpha_{ij}$, the charge placed in such material will
create a radial electric field, but due to the magnetoelectric effect there
will also appear magnetization
\begin{equation}
M_i = \sum_j \alpha_{ij} E_j
\end{equation}
For the diagonal magnetoelectric tensor $\alpha_{ij}$ this magnetization
will also be radial and have the shape of an ellipsoid. By
the same Helmholtz decomposition (\ref{eq:deco}) we see that it will have a
divergence-full component (the spherical part of moments, shown in Fig.~\ref{FIG:11}) and a
divergence-free quadrupole-like component. And the first one is
equivalent to having magnetic monopole at the position of a probe
charge.

The existence of such monopoles can lead to several nontrivial
consequences, in particular in the transport properties of such
systems~\cite{Khomskii-MMUT}. One can also have measurable effects when one places and
moves charges above the surface of such magnetoelectrics. Image
charge created in this case inside the magnetoelectric would create a
magnetic monopole, the magnetic field of which outside the sample can be
measured experimentally. Such measurements indeed confirmed this
picture~\cite{Meier} --- probably the
latest paper in which I.~Dzyaloshinskii is a coauthor. I suppose this
story will be described in details in another article in this volume,
by N.~A.~Spaldin.

\bigskip
\hbox to\hsize{\hfill \vrule width3cm height0.33pt\hfill}
\bigskip

In conclusion, we can say that there exist indeed diverse and very
interesting electric effects at different magnetic textures. This is
yet another manifestation of the nontrivial interplay of magnetic
and electric degrees of freedom in solids --- the field pioneered by
Dzyaloshinskii more that 50 years ago, and which still produces more and more new
surprises.

\section*{Acknowledgements}
I am grateful to many people with whom I collaborated over many
years on the topics discussed in this article. This work was
supported by the Deutsche Forschungsgemeinschaft (DFG, German
Research Foundation) -- Project number 277146847 -- CRC 1238

%\catcode`\@=11
%i\show\thebibliography

\end{document}